\documentclass{aa}  

\usepackage{graphicx}
\usepackage[dvipsnames]{xcolor}

\usepackage{txfonts}
\usepackage{lipsum}
\usepackage{subcaption}         
\usepackage{lscape}             
\usepackage{placeins}           

\usepackage{float}
\usepackage{threeparttable}
\usepackage{url}
\usepackage[bookmarks, colorlinks, breaklinks, 
]{hyperref}  
\hypersetup{linkcolor=blue,citecolor=MidnightBlue,filecolor=black,urlcolor=MidnightBlue}

\newcommand{\planck}{{\em Planck\,}}
\newcommand{\xmm}{{XMM-\textit{Newton}\,}}
\newcommand{\chandra}{{\em Chandra\,}}

\begin{document}

   \title{\planck ~Constraints on Turbulence in the Coma Cluster}



   \author{B. Sigal\inst{1,2}\thanks{baptiste.sigal@utoulouse.fr},
   E. Pointecouteau\inst{1},
   N. Clerc\inst{1},
   S. Dupourqué\inst{1},
   A. Molin\inst{1},
   F. Mernier\inst{1},
   R. Adam\inst{3},
   T. Dusserre\inst{4,2},
   F. Pajot\inst{1}
   }
    \authorrunning{B. Sigal et al.}
    \titlerunning{Turbulence in the Coma cluster}

   \institute{Univ Toulouse, CNES, CNRS, IRAP, Toulouse, France
                \label{IRAP} 
            \and Centre National d’Etudes Spatiales (CNES), Toulouse, France
            \and Université Côte d’Azur, Observatoire de la Côte d’Azur, CNRS, Laboratoire Lagrange, France
            \and Université Paris-Saclay, CNRS, Institut d’Astrophysique Spatiale, 91405 Orsay, France
            }

   \date{Received 22 May 2026 / Accepted 27 July 2026}

  \abstract
    {Turbulence within the intracluster medium (ICM)  influences galaxy cluster thermodynamics and virialisation, contributing to non-thermal pressure support and impacting hydrostatic mass estimates. Characterising this turbulence through thermodynamic fluctuations remains observationally challenging due to the non-linear relationships between observables such as density and pressure, and the underlying velocity field.}
    {This study aims to constrain the properties of ICM turbulence by performing a comprehensive reanalysis of the Sunyaev-Zel'dovich (SZ) surface brightness fluctuations based on \planck\ survey observations of the Coma cluster.}
    {We analyse the 2D power spectrum of SZ fluctuations, modelling the underlying 3D pressure fluctuation power spectrum assuming Kolmogorov-type turbulence. We infer key parameters from  a simulation-based inference framework relying on normalizing flows to accurately recover posterior distributions.}
    {By constraining the pressure fluctuation power spectrum, we are able to infer the properties of turbulence in the Coma cluster,
    finding a large injection scale of $l_{\text{inj}} = 540^{+450}_{-200}$ kpc, a slope of $\alpha = 3.50_{-0.46}^{+0.50}$ (under Gaussian prior), and a substantial 3D Mach number of $\mathcal{M}_{3D} = 0.60^{+0.13}_{-0.09}$. These values correspond to turbulent velocities in the range $\sigma_{v,\text{ }3D} = 357-1095$~km/s and a non thermal pressure fraction of $P_{\text{turb}}/P_{\text{tot}} = 0.17_{-0.04}^{+0.06}$.}
    {Our results are consistent with recent direct velocity measurements from XRISM, supporting a scenario of significant turbulence in the Coma cluster and highlighting the complex interplay of dynamical processes within its ICM. Our simulation-based inference approach applied to SZ fluctuations paves the way for systematic multi-probe studies combining SZ and X-ray data, as well as direct and indirect observations.}

   \keywords{Turbulence -- 
                Galaxies: clusters: intracluster medium --
                Galaxies: clusters: individual: Coma --
                Methods: statistical
               }

   \maketitle
   \nolinenumbers


\section{Introduction}
\label{s:int}
The intracluster medium (ICM) represents a complex, multi-phase plasma whose dynamics play a crucial role in the assembly and evolution of large-scale cosmic structures \citep{Sarazin1988,McNamara2007,Fabian2012,Kravtsov2012,Planelles2015}.  Gas motions in the ICM arise from a combination of hierarchical merging events \citep{Zuhone2022}, continuous accretion from the cosmic web \citep{Dolag2005, Vazza2017}, core sloshing phenomena \citep{Markevitch2007, Roediger2012, ZuHone2013}, and active galactic nucleus (AGN) feedback \citep{McNamara2012,Hlavacek-Larrondo2022}. 

The ICM gas dynamics sustain bulk and turbulent motions that produce a non-thermal pressure support, expected to account for up to $\sim 20-30$\% of the total pressure budget, with the fraction increasing towards cluster outskirts where accretion shocks and merger-induced motions dominate \citep{Dolag2005,Nelson2014, Vazza2018, Ayromlou2024}. 
The assumption of hydrostatic equilibrium, which underpin mass estimation techniques based on X-ray \citep{Ettori2013, Arnaud2010} or Sunyaev-Zel'dovich observations \citep{SZ1972} may not hold due to the presence this non-thermal component.

Consequently, it introduces systematic biases in the cluster mass estimates by 5-20\%, depending on dynamical state and radius \citep{Biffi2016,Pratt2019, Angelinelli2020}, that are key in cosmological analyses using the cluster population as probe \citep{Vikhlinin2009, PlanckCollaboration2016, Mantz2022,Clerc2023}. 

Measuring and characterising turbulent motions in the ICM can be tackled through direct or indirect approaches \citep{Simionescu2019}.
Direct measurements using the characterisation of X-ray spectral lines centroid shift and broadening have recently been possible from the advent of spatially resolved high-resolution X-ray spectroscopy with the \textit{Hitomi} satellite \citep{HitomiMission2018} and its successor, the XRISM satellite \citep{XRISMmission2025}. Their observations of several clusters have unveiled surprisingly low levels of measured turbulence at the centre of clusters \citep{Hitomi2016, Hitomi2018, XrismCollaboration2025a, XrismCollaboration2025b, XrismCollaboration2025c, XrismCollaboration2025d}. These velocity and velocity dispersion measurements, restricted to limited regions of the central part of clusters due to the modest spatial resolution of \textit{Hitomi} and XRISM,  convey a limited view of the whole turbulent velocity field at play in the ICM.

Indirect approaches examine the statistical properties of thermodynamic quantities through fluctuations in X-ray and Sunyaev-Zel'dovich (SZ) emissions. These studies link them to the ICM fluctuations in temperature \citep{ Hofmann2016}, electron density \citep[e.g.][]{Churazov_2012}, and pressure \citep{Schuecker2004, Romero2024, Adam_2025}, assuming a theoretical scaling relation between the turbulent velocity power spectrum and the power spectra of these thermodynamic quantities \citep{Gaspari2014, Simonte2022, Zhuravleva2023}. SZ fluctuations directly trace pressure perturbations, and their linear density dependence (in contrast to the quadratic scaling of X-ray emission) makes them a more sensitive probe at larger physical scales. While fluctuation analyses currently provide access to more spatial scales, their interpretation in terms of pure turbulence is complicated by several interfering factors  such as large-scale bulk motions from mergers or gas inflows, gas clumping from unresolved substructures, projection effects along the line of sight \citep{Nagai2011,Gaspari2014}. These processes not only contaminate velocity interpretations but can also actively perturb the local thermodynamics, thereby generating spurious fluctuations indistinguishable from turbulent signals. \\

The Coma cluster's proximity and extent\footnote{$R_{500}$ is the radius within which the mean matter density is 500 times the critical density of the Universe at the redshift of Coma.}  make it an ideal test case for turbulence investigation, either through direct or indirect measurements.
In this work, we adopt the values reported by \citet{PlanckCollaborationComa2013}, namely $R_{500} = 1.31$~Mpc and $z = 0.023$.
Coma has long been known as a dynamically disturbed cluster, its most prominent feature being the ongoing interaction with the infalling group NGC~4839, a view also completed by the presence of cold front, shocks and pressure edges seen from X-ray to radio wavelengths \citep{PlanckCollaborationComa2013, Sanders2020, Mirakhor2020, Bonafede2022, Churazov2021, Churazov2023}. \\

This cluster  has been extensively studied through multiple approaches.  The initial temperature fluctuation analysis using ROSAT data \citep{Schuecker2004} was followed by more detailed studies with \xmm  and \chandra\ observations \citep{Churazov_2012, Gaspari2013, Zhuravleva_2019}. These X-ray analyses have consistently revealed subsonic turbulent motions ($r < R_{500}$) with characteristic three-dimensional Mach numbers in the range $\mathcal{M}_{3D} \sim 0.2-0.8$ and energy injection scales of 200-500 kpc. The first SZ study by \citet{Khatri_2016} suggested a similar injection scale of approximately 500~kpc when qualitatively combined with \chandra constraints. 

These results require careful comparison with recent constraints from XRISM X-ray spectroscopy, which reported a poorly constrained injection scale, around $\sim 1000-2000$~kpc, and slope of the turbulent cascade, as well as measurements of the 3D Mach number under tension. \citet{xrismcollaboration2025xrismforecastcomacluster} reported $\mathcal{M}_{3D} = 0.24 \pm 0.015$, while \citet{Eckert_2025} obtained $\mathcal{M}_{3D} \approx 0.72^{+0.28}_{-0.22}$, the two studies being based on 2 and 3 XRISM pointings of $3\times 3$~arcmin$^2$, respectively. Such discrepancies underscore the fundamental challenges when interpreting velocity measurements during cluster merger phases, where bulk motions may dominate over turbulence. In such cases, this may potentially lead to an overestimation of turbulent parameters if bulk flows are misinterpreted as turbulent motions \citep{Donnert2018, Zhang2026}. \\

In this paper, we revisit the case of Coma's ICM turbulence, from a thorough and quantitative analysis of its SZ signal fluctuations as seen by the \planck\ mission \citep{PlanckCollaboration2011,PlanckCollaboration2018} in the light of recent methodology evolutions with simulation based inference framework based on normalising flows \citep{Dupourque2023b}. In the next sections, we first present the SZ data used, then our methods to characterise the statistics of the SZ signal fluctuations through their 2D power spectrum. Our method for modelling the 3D pressure fluctuations, hence the understated turbulence power spectrum, into our observable is presented in section~\ref{s:tpb}. In Sec.~\ref{s:constraints}, we state our constraints on the ICM turbulence of Coma, and discuss them in Sec.~\ref{s:discu}.

We adopt a $\Lambda$CDM cosmology with $H_0 = 70$ km s$^{-1}$ Mpc$^{-1}$, $\Omega_M = 0.3$, and $\Omega_\Lambda = 0.7$ throughout this work. At the  redshift of Coma, $z=0.023$, 1~arcmin corresponds to 27.9~kpc. Hereafter, $\log$ denotes the base-10 logarithm and $\ln$ the natural logarithm.


\section{Data}
\label{s:dat}

This study is based on observations by the \planck\ mission \citep{PlanckCollaboration2011,PlanckCollaboration2018} at millimetre wavelengths. We utilise the SZ-reconstructed all-sky Comptonisation parameter ($y$) map obtained through the Modified Internal Linear Combination Algorithm (MILCA) \citep{milca2013,planck-ymap2016}, which combines the six HFI frequency channels (100-857 GHz). In this work, we used the non-public MILCA map with an effective angular resolution of 7~arcmin FWHM.

The Coma cluster region was extracted from the \planck\ $y$-map in HEALPix format \citep{Gorski_2005} using a tangent plane equatorial projection. The extracted patch covers $20 \times R_{500}$ on each side, over a 1024$\times$1024 pixel grid centred on the cluster coordinates taken from the \planck SZ catalogue, PSZ2 \citep{PSZ2}: $\alpha = 194.9118^{\circ}$, $\delta = 27.9537^{\circ}$. $R_\textrm{500} = 1.31$~Mpc is taken from \citep{PlanckCollaborationComa2013} and corresponds to an angular extent of $\theta_\textrm{500}=47$~arcmin. The reprojected pixel scale of $\sim$57 arcsec oversamples the native HEALPix resolution of $\sim$1.72 arcmin pixels for $N_{\rm side}=2048$. Following \citet{PIPV2013}, we preserved the pixel to pixel correlations induced by the reprojection process and HEALPix pixel oversampling to ensure accurate propagation of uncertainties and induced correlation in subsequent analyses.


\section{Statistics of the SZ brightness fluctuations}
\label{s:isz}

Our investigation into the fluctuations of the SZ signal is conducted using the \planck\ $y$-map, with these fluctuations expressed in terms of the dimensionless Comptonisation parameter $y$.
We recall that the SZ effect arises from the inverse Compton scattering of CMB photons by hot electrons in the intra-cluster medium (\citealt{SZ1972}, and \citealt{SZ} for a recent review). The intensity of this effect is directly proportional to the integrated pressure along the line of sight:

\begin{equation}
y = \frac{\sigma_T}{m_e c^2} \int P_e \, dl,
\label{eq:ysz}
\end{equation}
where $\sigma_T$ is the Thomson cross-section, $m_e$ the electron mass, and $P_e$ the electron pressure. 

The statistical properties of the SZ fluctuations are encoded in their two-dimensional power spectrum (i.e., the power spectrum of the $y$-map). We use this observable to characterise the SZ fluctuations.

\subsection{2D power spectrum}
\label{s:pws}

The two-dimensional power spectra are computed using the Mexican hat filtering method described in \citet{Arevalo_2012}, and applied to the X-COP sample in X-rays by \citet{Dupourque2023b}. This method provides an estimation of the power spectrum through filtering an input image on multiple scales.

The power at a given spatial frequency $k$ is derived from the variance of the image after convolution with the Mexican hat filter. This approach not only captures the spatial distribution of fluctuations of scale $k$, but also accounts for masked or incomplete regions in the data. The latter capability is especially valuable in our analysis, as it allows us to robustly estimate the power spectrum even in the presence of masked sources or regions.

By applying this method across a range of spatial frequencies $k$, we obtain a discrete estimation of the full power spectrum of the image. For details on the formalism, including the computation of the power spectrum and error propagation, we direct the reader to App. A of \cite{Arevalo_2012} and to Eqs~(F.5), (F.8), and (F.9) in \citet{Dupourque2023b}. 
This method introduces a moderate bias, which can be corrected following App. B of \cite{Arevalo_2012}. We implemented this correction assuming a constant spectral slope. As the residual bias remains consistent across the analysis, it does not affect our final results.

\subsection{Power spectrum of the noise}
\label{s:noi}

{We first applied the power spectrum estimation method to characterise the statistics of noise in the \planck\ SZ map. We assumed that the statistical properties of the noise are constant over the whole image. We considered only the outer regions, i.e., $r>5\times R_\textrm{500}$, masking the inner parts to avoid any contamination by the Coma SZ signal. The derived power spectrum thus accounts for the instrumental noise, systematics and the residual astrophysical signals which may have leaked from components other than SZ in the $y$-map ILC reconstruction  (e.g., unresolved submillimetre and radio point sources, dust or Galactic synchrotron emissions).} 

The power spectrum of the noise over the Coma map, $\mathcal{P}_N$, is then used to draw random realisations of the noise,  accounting for the  oversampling of the HEALPix pixels in the tangential reprojection and assuming an inhomogeneous correlated Gaussian noise as described in \citet{PIPV2013}. Individual noise simulations, expressed as $\mathcal{N}_{sim}(\textbf{r}) = \mathcal{FT}^{-1}[\mathcal{FT}[W]\sqrt{\mathcal{P}_N}](\textbf{r})$ where $W(\textbf{r})$ is a white noise image realisation, then multiplied by the square root of the noise power spectrum.
From $n=30,000$ noise simulations we can estimate the covariance matrix of the input SZ $y$-map for Coma, through $\mathbf{C}=\mathbf{N}_{sim}^T \mathbf{N}_{sim}$, where $\mathbf{N}_{sim}$ is a $n\times m$ matrix encompassing all noise map simulations, with $m$ the number of pixels in the map.
In the following analysis, we focus on the central region of our patch, specifically a square area of $4R_{500} \times 4R_{500}$, which corresponds to 205~pixels on each side, thus $m=205^2$ pixels.

\begin{figure}[!t]
    \centering
    \includegraphics[width=0.9\linewidth]{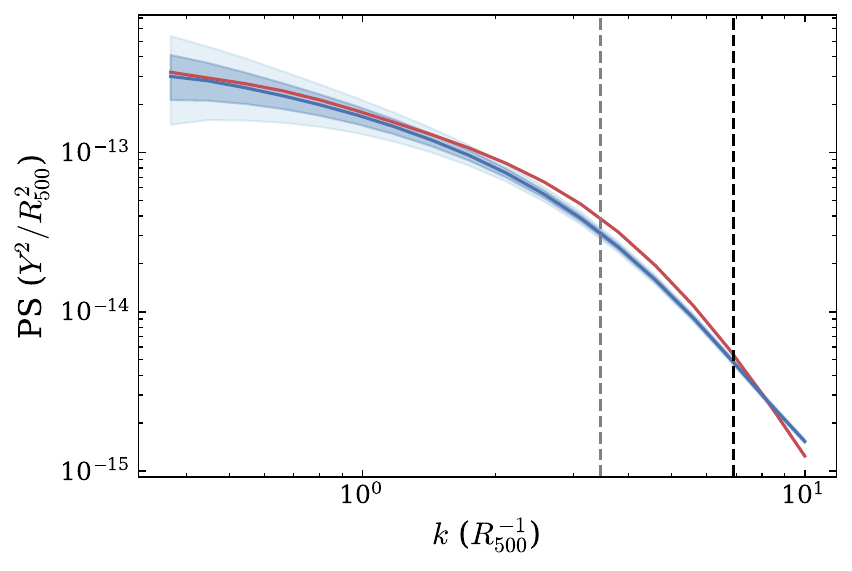}
    \caption{
    Comparison of the power spectrum of the measured noise (red line) with the median spectrum derived from 30,000 simulated noise realisations (blue). The dark and light blue shaded regions represent the 68\% and 95\% confidence envelopes around the median, respectively. The vertical dashed black and grey lines mark the scales corresponding to one and two times the \planck\ PSF of 7~arcmin FWHM resolution, respectively.
}    
\label{f:noi}
\end{figure}

To validate the consistency between the simulated noise map and the observed noise, we compared their respective power spectra, as illustrated in Fig.~\ref{f:noi}. This comparison ensures that the statistical properties of the simulated noise accurately reproduce those of the actual observational data. Small-scale differences arise from the Mexican hat bias \citep[App. B]{Arevalo_2012}, as discussed in the previous section, but these do not affect our analysis since they impact modes below our conservative cut at twice the \planck\ PSF.

\begin{figure*}[t]
    \centering
    \includegraphics[width=0.3\linewidth]{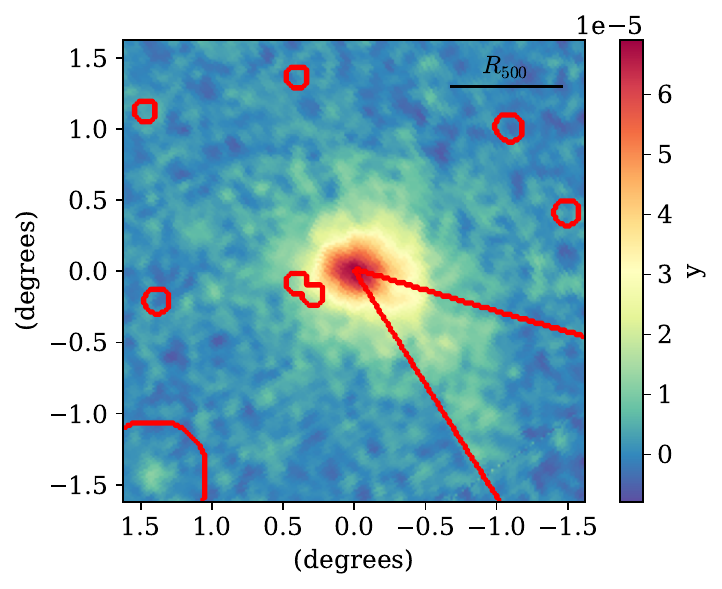}
    \includegraphics[width=0.3\linewidth]{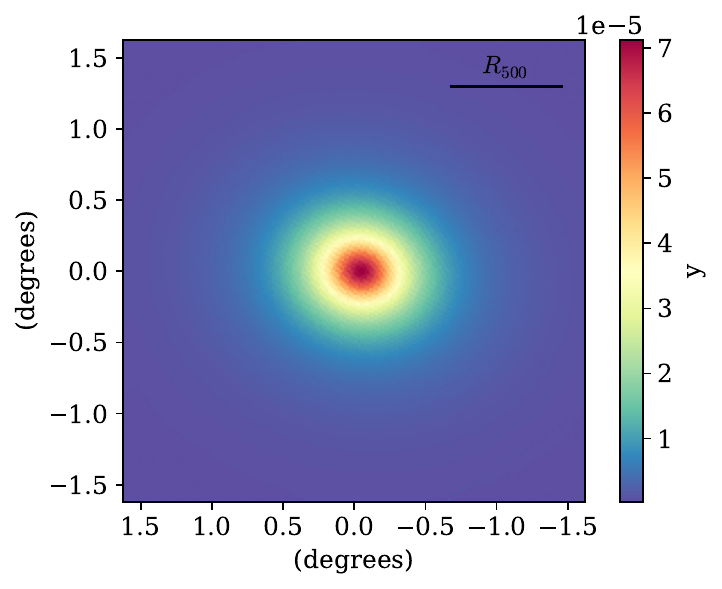}
    \includegraphics[width=0.31\linewidth]{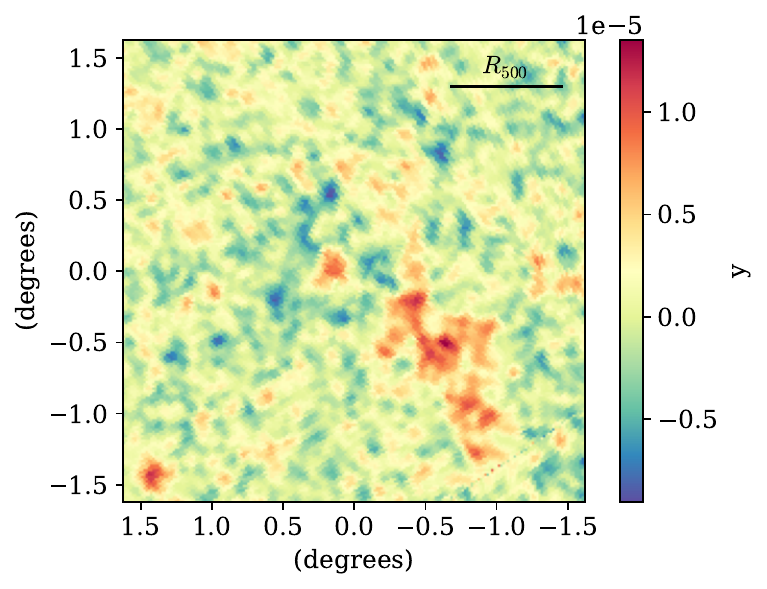}
    \caption{\textit{(left)} The \planck\ Coma cluster $y$-map over a patch of $4 R_\textrm{500}\times 4 R_\textrm{500}$ centred on the PSZ2 catalogue coordinates, i.e., $\alpha = 194.9118^{\circ}$, $\delta = 27.9537^{\circ}$. The red circles indicate the positions of masked known point sources (see Section~\ref{s:gnfw}). \textit{(centre)} Best fit model. \textit{(right)} Residual SZ fluctuation map.}
    \label{f:szb}
\end{figure*}

\subsection{Power spectrum of SZ fluctuations}
\label{s:msz}

\subsubsection{Fluctuation of the $y$-map}
\label{s:gnfw}

To extract the map of $y$ fluctuations, we first needed to subtract the main SZ emission. The mean emission is modelled using a generalised Navarro-Frenk-White profile \citep[gNFW,][]{Nagai2007,Arnaud2010}, which provides an analytical representation of the radial distribution of the ICM thermal pressure arising from the gas infall within the  dark matter dominated  gravitational potential:

\begin{equation}
    P(x) = \frac{P_0\times P_{500}}{(c_{500}x)^\gamma[1 + (c_{500}x)^\alpha]^{\frac{\beta - \gamma}{\alpha}}}
    \label{e:gnfw}
\end{equation}

\noindent with $P_0$ the normalisation, $P_{500}$ the pressure ($keV/cm^3$) at a density contrast of 500, $c_{500}$ the inflection point around radius $r_s$ with $r_s=R_{500}/c_{500}$, $\gamma$ the central slope ($r \ll r_s$), $\beta$ the outer slope ($r \gg r_s$), $\alpha$ modulates how smoothly the slope changes from $\gamma$ to $\beta$ around $r_s$, $x = r/R_{500}$ with $r$ the radial distance from the centre. 
The gNFW radial model is commonly used for spherically symmetric clusters.
To accurately capture the intrinsic $y$ fluctuations about the mean emission of the Coma cluster, we allow the 2D projection of the gNFW profile onto the plane of the sky to be elliptical.
This makes the model more versatile, without over- or under-fitting the 2D $y$ signal. For a detailed discussion on the impact of the model choice, we refer the reader to \citet{Zhuravleva2023} and \citet{Dupourque2023b}. Hereafter, we strictly applied the methodology described by these authors.

The gNFW profile is thus projected along the line of sight as a 2D ellipsoidal model on the plane of the sky, introducing two additional parameters: the ellipticity, $e$, and the orientation of the ellipse, $\theta$, which is the angle between its major axis and the east direction, increasing northward. The resulting projected pressure model map is then converted into a $y$-map using Eq~\ref{eq:ysz}.

When fitting the main emission model, the following parameters were left free: the cluster centre position ($x_c$,$y_c$), $P_0$, $c_{500}$, $\beta$, $\theta$ and $e$. $\alpha$ and $\gamma$ are fixed to 1.33 and 0, respectively, following the choice of parameterisation adopted for the X-COP sample \cite{Ghirardini_2019}. The clusters of this sample are indeed local and massive clusters and therefore analogous to Coma. This choice is further justified by the moderate spatial resolution of the \planck\ $y$-map, which prevents us from constraining the central regions.

~\\
The fit of the mean SZ signal model was performed using a gradient-based optimisation method \footnote{The Levenberg–Marquardt algorithm implemented in \texttt{curvefit} from the Python package $SciPy$}. The fit was applied to the \planck\ Coma image, cropped to a $4R_{500} \times 4R_{500}$ patch centred on the cluster (Fig.~\ref{f:szb}).
Known point sources from the \planck\ Catalogue of Compact sources \citep{pccs}  and clusters from the PSZ2 catalogue \citep{PSZ2} have been masked (see left panel in Fig.~\ref{f:szb}.)

To avoid bias from the emission related to NGC~4839's infall onto Coma (see Sec.~\ref{s:int}), we applied a pie-slice mask excluding both the subgroup and its tail during the gNFW mean model fitting.
The pie-slice mask (defined from SAO ds9) shown in Fig.~\ref{f:szb} is centred on Coma's PSZ2 position, starting at $125^\circ$, measured northward from east, and with an angular extent of $40^\circ$.

The best fit parameters are provided in Table~\ref{table:1} and the best fit model is shown in the middle panel of Fig.~\ref{f:szb}.  The  position of the centre, $(x_c,y_c)$, is fitted in pixels. The $y$ fluctuation map is obtained by subtracting the best fit mean model from the $y$-map (see right panel of Fig~\ref{f:szb}).

\begin{table}[b]
\caption{Best fit parameters for the mean pressure model}
\label{table:1}    
\centering                        
\renewcommand{\arraystretch}{1.2}
\begin{tabular}{c c}      
\hline\hline               
  Parameters & Values \\
\hline
  $x_c$ (RA)  & $194.903 \pm 0.002$\\
  $y_c$ (DEC)  & $28.009 \pm 0.002$\\
  $\theta$ ($^\circ$) & $12.2 \pm 1.4$\\
  $e$ (-) & $0.40 \pm 0.01$\\
  $P_0$ (-) & $9.59 \pm 0.18$\\
  $c_{500}$ (-) & $1.86 \pm 0.05$\\
  $\beta$ (-) & $4.68 \pm 0.08$ \\
\hline                      
\end{tabular}
\end{table}

\subsubsection{SZ power spectrum}
\label{s:szp}

\begin{figure*}[t]  
    \centering
    \includegraphics[width=0.4\linewidth]{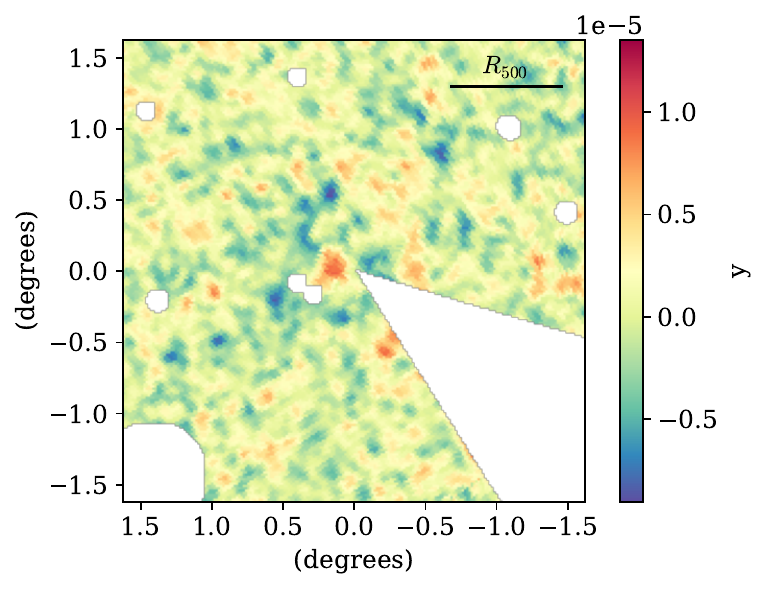}
        \includegraphics[width=0.43\linewidth]{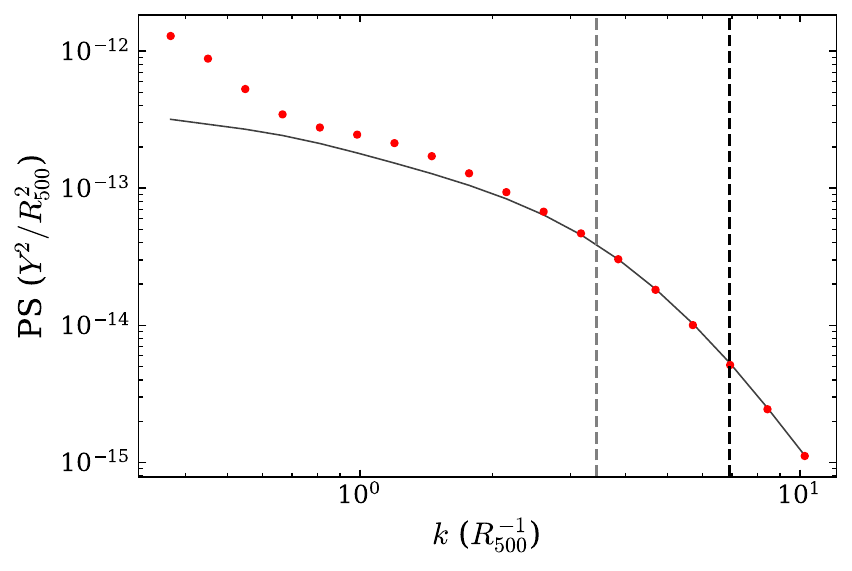}
    \caption{\textit{(left)} SZ fluctuation map. The known point sources and known clusters from \planck catalogues are masked. The South-Eastern region of NGC~4839 is masked by a pie-slice mask. \textit{(right)} Power spectrum of the SZ $y$ fluctuations (red points) compared to the noise power spectrum (gray line). The uncertainties on the best mean SZ model (see Table~\ref{table:1}), though propagated, are tiny. The associated error bars are much smaller than the dots.
    The vertical dashed black and grey lines mark the scales corresponding to one and two times the \planck\ PSF of 7~arcmin FWHM resolution, respectively.}
    \label{f:psz}
\end{figure*}

We derived the SZ signal power spectrum by applying the methodology outlined in Section~\ref{s:pws} to the $y$ fluctuation map, after masking both point sources and the NGC~4839 pie-slice sector. The resulting 2D power spectrum, $\hat{\mathcal{P}}_{2D}(k)$, was computed across 18 $k$-modes, linearly spaced in logarithmic scale from $k_{\rm min} = 1/(2\sqrt{2}R_{500})$ to $k_{\rm max} = 1/(0.1R_{500})$, corresponding to spatial scales from 3.7~Mpc to 131~kpc, respectively. Figure~\ref{f:psz} presents the masked fluctuation map and  derived power spectrum.

The shape of the recovered power spectrum may be impacted by the  mask definition, particularly in our case by the inclusion or exclusion of the NGC~4839 region. This point is discussed in Section~\ref{s:mas}. Additionally, the choice of centre for the 2D elliptical model may also affect the power spectrum computation. We investigate this impact in Sec.~\ref{s:cen}.

\section{Pressure fluctuations in the ICM}
\label{s:tpb}
As shown in Eq.~\ref{eq:ysz}, fluctuations in the SZ Comptonisation parameter trace pressure variations within the Coma cluster's ICM,  as X-ray surface brightness fluctuations trace density perturbations \citep[e.g.][]{Churazov_2012,Simionescu_2019}.

These pressure fluctuations arise from multiple physical processes involving gas motions that displace matter and perturb the local thermodynamic equilibrium. In our analysis of the Coma cluster, the dominant contributing processes include turbulent motions generated by hierarchical mergers and continuous accretion \citep{Gaspari2014,Donnert2018} or compression waves from accretion shocks at the cluster outskirts \citep{Brunetti2014,Vazza2017}, as well as large-scale bulk flows from ongoing mergers or gas sloshing \citep{Markevitch2007,ZuHone2016}. However, the angular resolution of the \planck\ $y$-map ($\sim$7 arcmin) effectively prevents us from resolving the smaller-scale influence of AGN feedback from jets and buoyantly rising bubbles \citep{McNamara2007,Hlavacek-Larrondo2015}.

\subsection{The power spectrum of pressure fluctuations}
\label{s:3dpws}

In order to model the observed power spectrum of SZ signal fluctuations  derived in the previous section, we considered that it arises from the pressure fluctuations in the ICM of the Coma cluster. We assume these fluctuations follow a Gaussian random field with a power spectrum that can be modelled as a single power law with slope $\alpha$, suppressed at small and large scales (large and small spatial scales) with two decreasing exponentials parameterised according to two characteristic cutting scales that we note $k_{inj}$ and $k_{dis}$, respectively:

\begin{equation}
    \bar{\mathcal{P}}_{3D}(k) = \sigma^2 \frac{e^{-(k/k_{dis})^2}e^{-(k_{inj}/k)^2}k^{-\alpha}}{\int4\pi k^2dk e^{-(k/k_{dis})^2}e^{-(k_{inj}/k)^2}k^{-\alpha}}
    \label{e:p3d}
\end{equation}

\noindent where $\sigma$ is the normalisation of the power spectrum. The wavenumber $k$ is related to the spatial scale, $r$, by $k = 1/r$. $\bar{\mathcal{P}}_{3D}(k)$ denotes the 3D power spectrum of the pressure fluctuations, which should be distinguished from a single realisation of the pressure fluctuation considered as a Gaussian random field.

With this formulation, we interpret all pressure fluctuations as turbulence (see Sec.~\ref{s:tur}), following the approach and formalism developed by \citet{Dupourque2023b} for the analysis of fluctuations of X-ray surface brightness and hence gas density. In this framework, the adopted turbulence model is the simplest possible, that is a scale-free cascade with a slope $\alpha$ analogous to that of \citet[][, for which $\alpha=11/3$]{Kolmogorov1962} between two characteristic scales, the dissipation, $k_{dis}$, and the injection, $k_{inj}$.

\subsection{Modelling the 2D SZ power spectrum}

\subsubsection{A simulation-based inference approach}
The characterisation of pressure fluctuations in the ICM via the SZ signal demands a robust statistical framework to address the inherent stochasticity of the observable (hereafter we assume turbulence in the ICM as the source of the pressure fluctuations) and its associated error budget. A classical Bayesian  approach, which relies on an analytical likelihood function, is not applicable in this context, because of the complexity of the calculations this would involve. Moreover, we only have a unique observation of finite size as a realisation of the stochastic field. This introduces an additional source of variance. This effect dominates the error budget at larger spatial scales and must be rigorously modelled and accounted for to prevent the underestimation of uncertainties \citep{Clerc2019,Cucchetti2019}.

Simulation-based inference (SBI) offers a compelling solution by leveraging forward modelling to learn the likelihood function directly from synthetic observations. This approach enables the assessment of the full error budget, including sample variance and potential data correlations, by generating mock datasets that replicate the observed signal \citep{Dupourque2023b,Molin_2025}. By training a neural network to approximate the likelihood, we can sample the posterior distribution of the parameters governing the 3D power spectrum of pressure fluctuations (e.g., Eq.~\ref{e:p3d}). This methodology ensures that the stochastic nature of the observable is fully propagated into the parameter constraints as it naturally marginalises over the possible realisations of the random field, yielding more reliable and physically interpretable results. In our work, we have relied on the \texttt{sbi} python package \citep{sbi,boelts_sbi_2025}.

Our approach builds upon the methodology developed in \citet{Dupourque2023b}, where a similar framework was applied to X-ray surface brightness fluctuations, and extends it to the SZ signal. We also note that this method has been applied to the XRISM observations of the Coma cluster to parameterise the ICM turbulence from the measurements of X-ray spectral lines centroid shift and broadening \citep{Eckert_2025}.

\subsubsection{Modelling and training}
\label{s:model_train}

The 2D power spectrum of SZ signal fluctuation, $\hat{\mathcal{P}}_{2D}$, is the observable against which we want to test our model of 3D pressure fluctuation. To proceed, we strictly followed the method presented in \citet{Dupourque2023b}. We outline below the main steps and the adaptation for the SZ signal with respect to the X-rays:

\begin{itemize}
    \item Gaussian random field of pressure fluctuations.
    We generated random realisations of the pressure fluctuation field from the 3D power spectrum  given in Eq.~\ref{e:p3d} over a 3D grid covering $4R_\textrm{500}\times4R_\textrm{500}$ on the plane of the sky and $10R_\text{500}$ along the line of sight. We kept free three of the four parameters modelling  $\bar{\mathcal{P}}_{3D}(k)$: $\sigma$, $\alpha$, $l_{inj}$. Their values for each realisation are drawn from uniform priors (see below). We fixed $l_{dis}$ to $1$ kpc. The actual true dissipation scale is indeed expected to be way smaller than the smallest scale available in our 7~arcmin FWHM \planck\ map \citep{Zhuravleva_2019}. ~\\
    
    \item Mean gas pressure component.
    We generated mock pressure  distribution by drawing Gaussian realisations about our best fit model derived for the mean pressure and SZ signal of Coma (see Eq.~\ref{e:gnfw}), accounting for the statistical covariance between the parameters (as returned by the \emph{curvefit} fitting procedure -- see Table~\ref{table:1}).~\\
   
    \item Projection.
    The total 3D pressure model, including pressure fluctuations, is projected along the line of sight to produce a 2D elliptical $y$ model. We account for the resolution of the \planck\ $y$-map by convolving the projected map with a 7~arcmin FWHM Gaussian filter. ~\\
    
	\item Noise modelling.
    We added a realisation of the noise to each projected mock $y$-map, generated from our 2D power spectrum of noise computed in Sec.~\ref{s:noi}.~\\

    \item Subtraction of the 2D cluster model.
    The 2D mean model of the cluster (Table~\ref{table:1}) is subtracted from the projection to produce the SZ fluctuation map. ~\\

    \item 2D power spectrum of mock images.
    The 2D power spectrum of the SZ signal fluctuations, $\mathcal{P}_{\text{2D}}(k)$, is estimated exactly as the observed power spectrum, $\hat{\mathcal{P}_{2D}}$, as described in Sec.~\ref{s:pws}.
    ~\\
    
    \item Training hypothesis. The priors used for the three free parameters for the pressure power spectrum are uniform and such that $\log(l_{inj}) \sim \mathcal{U}(-0.60,0.40)$ with $l_{inj}$ in units of $R_{500}$, which corresponds to a physical scale $l_{inj} \in [0.25,2.5]\, R_{500}$, $\alpha \sim \mathcal{U}(2.5,4.5)$ and $log(\sigma) \sim \mathcal{U}(-0.6,0.28)$. Recent studies of Coma and other clusters, using direct and indirect measurements \citep{ Zhuravleva_2019,Dupourque2024,xrismcollaboration2025xrismforecastcomacluster, Adam_2025,Eckert_2025}, have shown that the slopes of the density, pressure and velocity power spectra are hard to determine from current observational constraints. To ease the convergence process, we also devise a training case involving Gaussian priors for the slope, $\alpha$, centred on the expected $11/3$ value of the pure hydrodynamical Kolmogorov turbulence spectrum with a standard deviation of $0.5$, that is $\alpha \sim \mathcal{N}(\mu_{\alpha}=11/3,\,\sigma_{\alpha}=0.5)$. Moreover, this is supported by dimensional analysis restraining $\alpha$ to values between 3 and 5 \citep{zhuravleva_constraints_2012}.
    ~\\
    
    \item Training simulations. We produced 200,000 mock fluctuation $y$-maps, this number has been used as a compromise between the large number of simulations required to correctly train SBI on sample variance, and the training time as well as the time required for producing the training set. They were used to train a neural network, specifically a {normalizing flow} based on a {masked autoregressive flow}, to approximate the likelihood function $p(\mathcal{P}_{\text{2D}} | \sigma, \alpha, l_{\text{inj}})$. We then employed {Sequential Neural Posterior Estimation (SNPE)} as the sampling algorithm to infer the posterior distribution of the parameters. This method is particularly adapted for fast inference on a single observation as in our case.
    
\end{itemize}

During the mock $y$-map simulation process, for computational efficiency we neglected the over-pixelisation arising from the reprojection of the \planck\ HEALPix $y$-map onto an equatorial tangential $y$-patch (see Sec.~\ref{s:dat}). This approximation does not bias our estimation of the 2D power spectrum for each simulated map, as the spatial scales affected lie below the intrinsic resolution of the $y$-map (i.e., 7~arcmin FWHM).
Given that our modelling and fitting of the 2D power spectrum are restricted to modes corresponding to spatial scales at least twice this size (see Fig.~\ref{f:psz}), any induced bias in the power spectrum remains fully negligible at the scales of interest. The 2D power spectrum (P2D) is computed at values of $k$ up to $2\sqrt{2} \, R_{500}$, i.e., at a maximum scale that is half the largest scale available in our map. This choice ensures sufficient statistics for the larger scales to cope with the impact of cosmic variance.

We also compared our 3D power spectrum model (Eq~\ref{e:p3d}) to the alternative power-law form without cut-offs proposed by \citet{Eckert_2025} (their Eq.~4).
Using identical parameters ($l_{\text{inj}}=900$~kpc, $\alpha=11/3$, $\log(\sigma)=-0.3$), we generated two sets of 1,000 simulations following the pipeline described above. The resulting 2D power spectra are statistically indistinguishable, with overlapping 68\% confidence intervals and a mean relative difference $\sim 9$\% , mainly at large scales ($>R_{500}$).
This demonstrates that the formulations are effectively equivalent in this regime and supports a consistent comparison with \citet{Eckert_2025}.

\begin{figure*}[h!]
    \centering
    \includegraphics[width=0.32\textwidth]{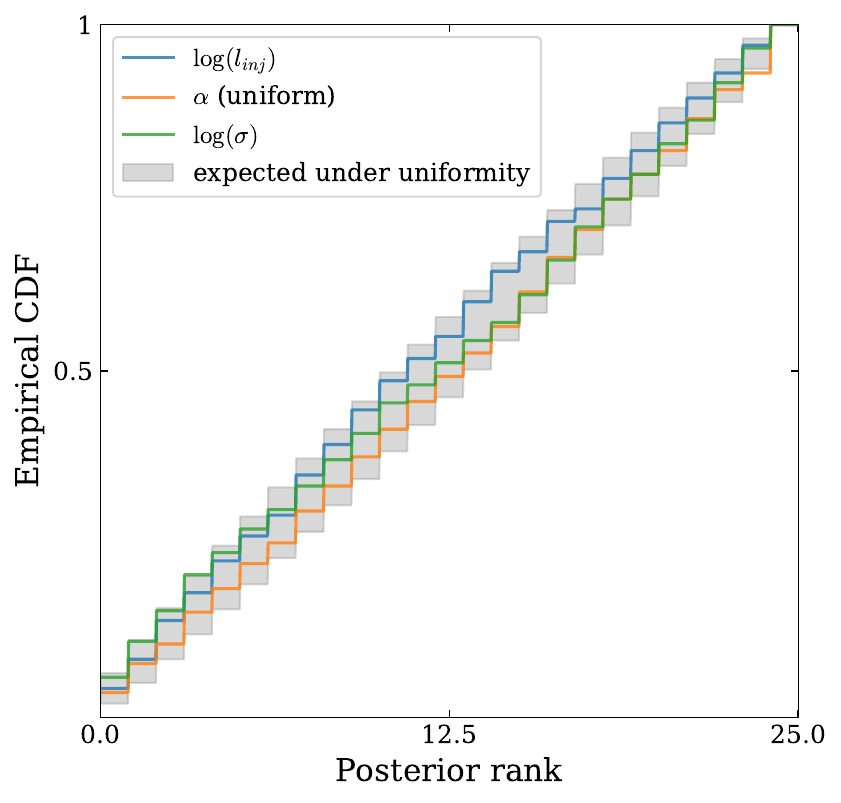}
    \includegraphics[width=0.32\textwidth]{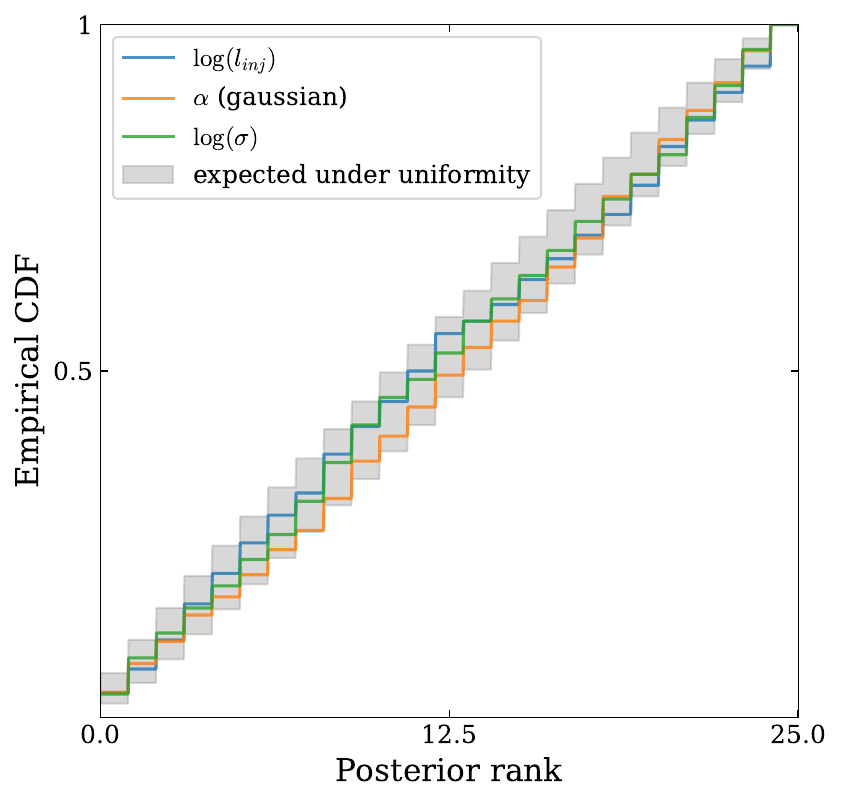}
    \includegraphics[width=0.32\textwidth]{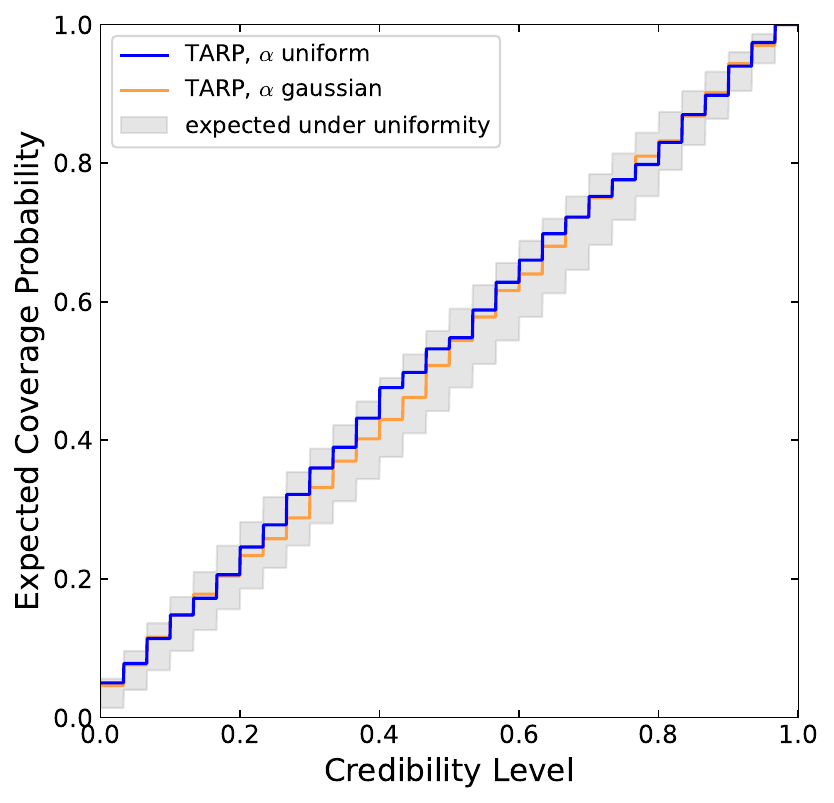}
     \caption{Training validation for our SBI approach. SBC results \textit{(left and middle panels)} and TARP diagnostics \textit{(right)} for models trained with uniform and Gaussian priors, respectively.}
      \label{f:val}
\end{figure*}

\subsection{Training validation}

To ensure the robustness and reliability of our neural network training, we conducted a rigorous validation of its predictive performance through the following testing procedures.

\subsubsection{Simulation-based Calibration (SBC) in SBI}  

We implemented the Simulation-Based Calibration (SBC) method \citep{Cook2006,Talts2018} as implemented in the \texttt{sbi} python pipeline. This diagnostic technique serves as a necessary (though not sufficient) condition to identify poorly trained networks across the entire space of the prior distributions of the parameters. The SBC procedure comprises four steps. First, a new turbulent parameter set $\theta^*$ sampled from the training priors is generated, with corresponding simulated 2D power spectra $\mathcal{P}_{2D}^*$. Here $\theta^*$ stands for $[\log{(l_{inj})}, \alpha, \log{(\sigma)}]$. Secondly, posterior inference on these simulations is derived using the trained neural network. Then, it associates each simulated spectrum $\mathcal{P}_{2D}^*[i]$ with a posterior distribution $\{\theta\}_i$ and a set of $s$ posterior samples $\{\theta\}_{i,s}$. Finally, it establishes the rank distribution by computing the rank for each parameter set $\theta^*_i$, counting the fraction of posterior samples $\{\theta\}_{i,s}$ falling below the true value.

The SBC diagnostic is natively implemented in the \texttt{sbi.diagnostics} Python module. The primary validation metric is the Cumulative Distribution Function (CDF) of the computed ranks. Uniform rank CDFs indicate proper calibration, while deviations reveal network deficiencies. 

\subsubsection{TARP calibration check}

We also employed the Two-sample Analysis of Recovered Parameters  diagnostic \citep[TARP,][]{Lemos2023} implemented in the \texttt{sbi.diagnostics} library, which provides both necessary and sufficient conditions for assessing the performances of neural posterior estimation. The TARP procedure can be described in four steps. First, a new turbulent parameter set $\theta^*$ sampled from the training priors is generated, with corresponding simulated 2D power spectra $\mathcal{P}_{2D}^*$. Then, posterior inference on each simulated $\left\{\mathcal{P}_{2D}^*\right\}_i$ is derived, yielding posterior samples $\{\theta_s\}$. Thirdly, random reference points $\theta_r$ from the prior distribution are selected. Finally, for each simulation, the relative distance metric
    $|\theta_s - \theta^*| < |\theta_r - \theta^*|$ is computed,
    counting how many posterior samples $\theta_s$ fall within the hyper-sphere centred on $\theta^*$ with radius $|\theta_r - \theta^*|$.
    
The diagnostic output is the Expected Coverage Probability (ECP) plotted against credibility levels \(c\%\). For a well-calibrated network, the curve follows the \(y = x\) line, indicating that the true parameters \(\theta^*\) fall within the \(c\%\) highest posterior density regions with probability \(c\%\). Deviations from this diagonal indicate miscalibration in the neural posterior. It should be noted that this tool does not provide any credible interval for the curve to lie on, only a p-value testing the hypothesis of uniform distribution for the relative distances. In our case, the obtained p-values did not invalidate this hypothesis and we added the same kind of interval as for SBC.

\subsubsection{Diagnostics of our trained neural networks}

We evaluate the calibration of our two trained neural networks - one with a uniform prior on the slope parameter $\alpha$ and another with a Gaussian prior - using both SBC and TARP diagnostics, with results presented in Figure~\ref{f:val}.

The SBC analysis  demonstrates that for both prior configurations (uniform and Gaussian), the empirical CDFs of all three parameters remain within the 95\% confidence intervals of a uniform distribution. This is further assessed by Kolmogorov-Smirnov tests which do not invalidate that the ranks come from uniform distributions (p > 0.05).

The TARP diagnostics show excellent agreement with the expected $y=x$ diagonal relationship. The slight deviations indicate a mild under-confidence by the neural network in specific parameter regimes. This could result in conservative and wider posteriors. Such behaviour is preferable in scientific contexts, where underestimating uncertainties (in the case of over-confident models) would be more problematic.

\subsection{From pressure fluctuations to turbulence} 
\label{s:tur}

We purposefully modelled the 3D pressure fluctuation power spectrum with a function analogous to the simplest turbulent power spectrum (see Eq.~\ref{e:p3d}). 
Our SBI network is trained to recover the large scale exponential cut-off analogous to the turbulent injection scale, the power spectrum slope characterising the inertial range of the turbulent cascade and the logarithm of the normalisation. The last is directly linked to the turbulent velocity dispersion along the line of sight, $\sigma_{v,\text{ }1D}$, when linking  pressure fluctuations to turbulent velocities. 

In practice, we followed the methodology outlined in \citet{Adam_2025} for this conversion and adopted the scaling relation proposed by \citet{Zhuravleva2023}. This relation converts the normalisation of the pressure fluctuation power spectrum into a 3D Mach number. Such scaling relations are calibrated using numerical simulations, which establish empirical links between the power spectrum of fluctuations of thermodynamic quantities (e.g., pressure and density) and the underlying turbulent velocity field in the ICM \citep{Gaspari2014, Simonte2022, Zhuravleva2023}.
We used the following scaling relation:

\begin{equation}
    \mathcal{M}_{3D} = \sqrt{3} \times \left(a + g \frac{\delta \xi}{\xi}\right),
\end{equation}
where $\delta \xi / \xi = 2\sqrt{2 \ln 2} / \log (\sigma_{\ln})$, and $\sigma_{\ln}$ is defined as:

\begin{equation}
    \sigma_{\ln} = \sqrt{\ln\left(\frac{1 + \sqrt{1 + 4 \sigma^2}}{2}\right)},
\end{equation}
with \(\sigma\) representing the normalisation of the 3D pressure fluctuation power spectrum as defined in Eq.~\ref{e:p3d}. \citet{Zhuravleva2023} provide a parameterisation of this scaling relation as a function of the cluster's dynamical state. For the Coma cluster, we adopted the "In-Between" case from \citet{Zhuravleva2023}, as its morphology matches the middle panel of their Fig.~2. We used the corresponding parameters from their Table~1, columns (i) and (j), i.e., \(a = 0.19\) and \(g = 0.36\).

The 3D Mach number  is directly related to the ratio of turbulent energy to thermal energy in the ICM. This allows us to connect the observed pressure fluctuations to the underlying turbulent motions. The ratio of turbulent to thermal energy is given by:
\begin{equation}
    \frac{E_{\text{turb}}}{E_{\text{therm}}} = \frac{1}{2} \gamma (\gamma - 1) \mathcal{M}_{3D}^2,
    \label{eq:eturbetherm}
\end{equation}
where \(\gamma = 5/3\) is the polytropic index for a monoatomic ideal gas. In this case, the ratio of turbulent to thermal energy is equivalent to the ratio of their corresponding pressures, i.e., \(E_{\text{turb}}/E_{\text{therm}} = P_{\text{turb}}/P_{\text{therm}}\).

The turbulent pressure contributes to the total pressure budget of the ICM, which impacts the hydrostatic equilibrium assumption. When turbulent pressure is not accounted for, it introduces a bias in the mass estimates, referred to as the turbulent mass bias, \(b_{\text{turb}}\). This bias is defined as the ratio of turbulent pressure to the total pressure in the ICM (\(P_{\text{turb}} + P_{\text{therm}}\)):

 \begin{equation}
    b_{turb}=\frac{P_{turb}}{P_{tot}}=\frac{P_{turb}}{P_{turb}+P_{therm}}=\frac{\mathcal{M}_{3D}^2\gamma}{\mathcal{M}_{3D}^2\gamma+3}
\end{equation}

The 3D Mach number is defined as the ratio of the turbulent velocity dispersion, \(\sigma_{v,\text{ }3D}\), to the speed of sound in the ICM, \(c_s\), as $\mathcal{M}_{3D} = \sigma_{v,\text{ }3D}/c_s$, with 

\begin{equation}
    c_s = \sqrt{\frac{\gamma k_B T_{gas}}{\mu_{gas}m_p}}
\end{equation}

\noindent with the polytropic index $\gamma = 5/3$, the mean molecular weight of the intracluster plasma $\mu_{gas} = 0.6$, the temperature of the gas $T_{gas}$, the Boltzmann constant $k_B$, and $m_p$ the proton mass.

In the next section, we estimate the average turbulent velocity dispersion from the 3D Mach number, using sound speeds derived from the temperature profile of \citet{Simionescu_2013}. With temperatures ranging from $\sim$8.5 keV in the core to $\sim$2 keV at $\sim$2 Mpc, the corresponding sound speed interval is $\sim$[700–1500] km/s.

\begin{figure*}[t]
    \centering
    \includegraphics[width=0.32\linewidth]{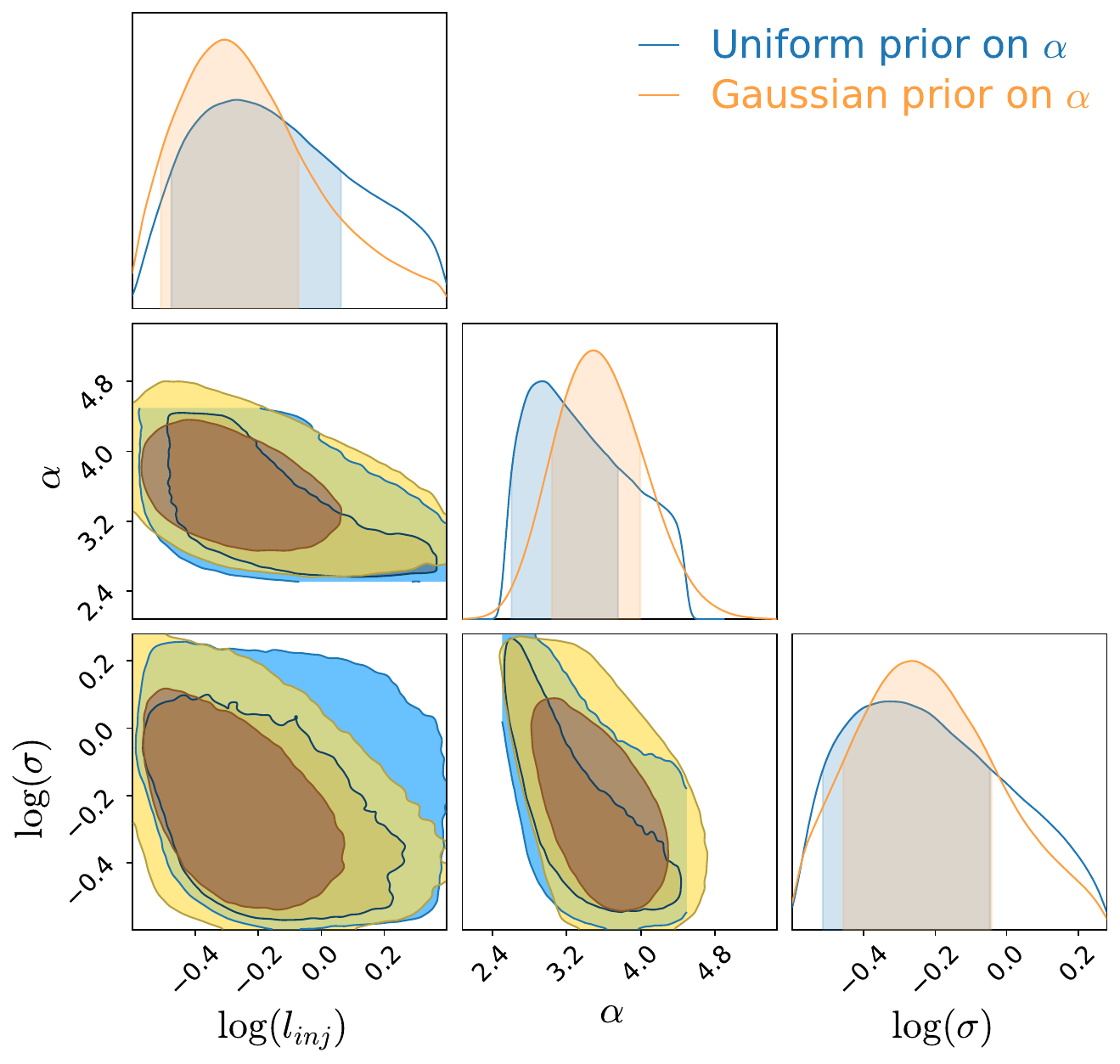}
    \hspace{0.05\linewidth}    \includegraphics[width=0.42\linewidth]{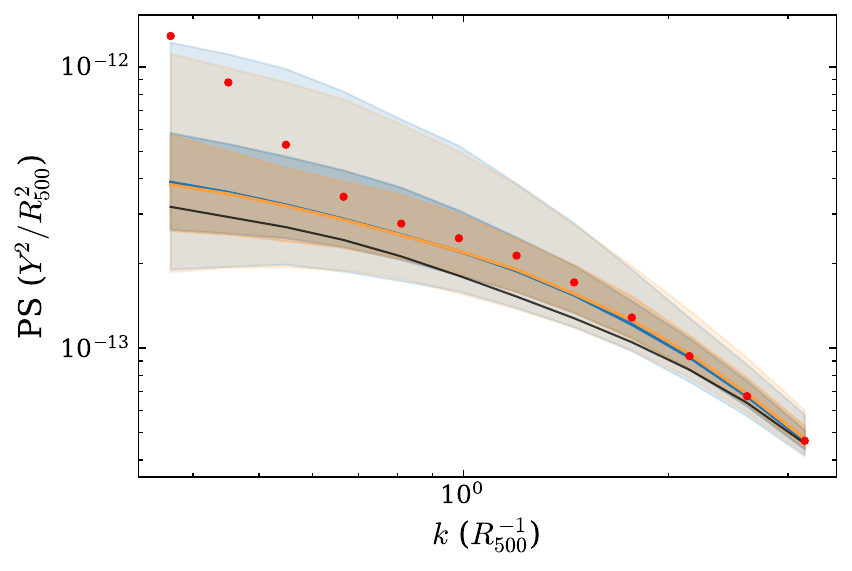}
    \caption{\textit{(left)} Posterior distributions on the three turbulent parameters obtained with SBI using SNPE. For this case, NGC~4839 was masked on the fluctuation map. \textit{(right)}  Posterior predictive check showing the measured 2D power spectrum of $y$ fluctuations (red points) compared to the median (with 68\% and 95\% confidence intervals envelop) over 500 posterior samples. On both panel the case of uniform and Gaussian priors are shown in blue and orange, respectively. The power spectrum of the noise is shown in black.}
    \label{f:post}
\end{figure*}

\section{Constraints on the turbulence in the ICM of Coma}
\label{s:constraints}

The posterior distributions inferred from our analysis for the turbulent parameters are shown in Fig.~\ref{f:post}. They are provided for the two sets of training priors for the power spectrum slope (uniform and Gaussian). The results highlight the impact of choice of priors on the derived constraints. Unsurprisingly, the use of a Gaussian prior on \(\alpha\) tightens the posteriors on all parameters, though without significant changes in the central values with respect to the posteriors derived for the uniform prior case. The right panel of Fig.~\ref{f:post} shows the measured 2D power spectrum compared to the median and associated dispersion of 500 posterior predictive realisations, demonstrating the goodness of fit.

The numerical results, detailed in Table~\ref{t:results}, reveal an injection scale of \(l_{\text{inj}} = 530^{+790}_{-200}\)~kpc (uniform prior case), consistent with turbulence driven by cluster-wide processes such as mergers and large-scale accretion. While the inferred values for the slope \(\alpha\) must be interpreted cautiously, they remain consistent with the Kolmogorov spectrum prediction of \(11/3\). Indeed, the uniform prior yields \(\alpha = 2.96^{+0.80}_{-0.35}\), whereas the Gaussian prior results in \(\alpha = 3.50^{+0.50}_{-0.46}\). The close agreement between these values confirms that the Gaussian prior does not significantly alter the estimate of the posterior's centre while providing a smooth regularisation. The inference of the two other parameters is robust against the choice of prior, with \(\log \sigma = -0.26^{+0.22}_{-0.20}\) and \(l_{\text{inj}} = 540^{+450}_{-200}\)~kpc in the Gaussian prior case.
Given the improved constraints on the slope and the consistent posterior distributions for both the injection scale and normalisation, we adopt the results obtained with the Gaussian prior as our reference results and use them in our subsequent analysis.

The inferred value of \(\log(\sigma)\) is physically interpreted using the relations established in the previous section. This yields a 3D Mach number of \(\mathcal{M}_{3D} = 0.60^{+0.13}_{-0.09}\), corresponding to an estimated 3D turbulent velocity dispersion of about \(\sigma_{v,\text{ }3D} \sim 357 - 1095\)~km/s (the range accounts for the variations of $c_s$ from the outer parts to the cluster centre, as explained in Sec. \ref{s:tur}). The turbulent-to-total pressure ratio, derived from the Mach number, is found to be \(0.17^{+0.06}_{-0.04}\), indicating a substantial contribution of turbulence to the ICM pressure budget. The derived values of $\mathcal{M}_{3D}$, $\sigma_{v,\text{ }3D}$, and $P_{\text{turb}}/P_{\text{tot}}$ remain fully consistent across the two other cases of scaling relations (``relaxed'' and ``unrelaxed'' prescriptions) provided by \citet{Zhuravleva2023}, thereby validating our intermediate case assumptions.

\begin{figure}[t]
    \centering
    \includegraphics[width=0.9\linewidth]{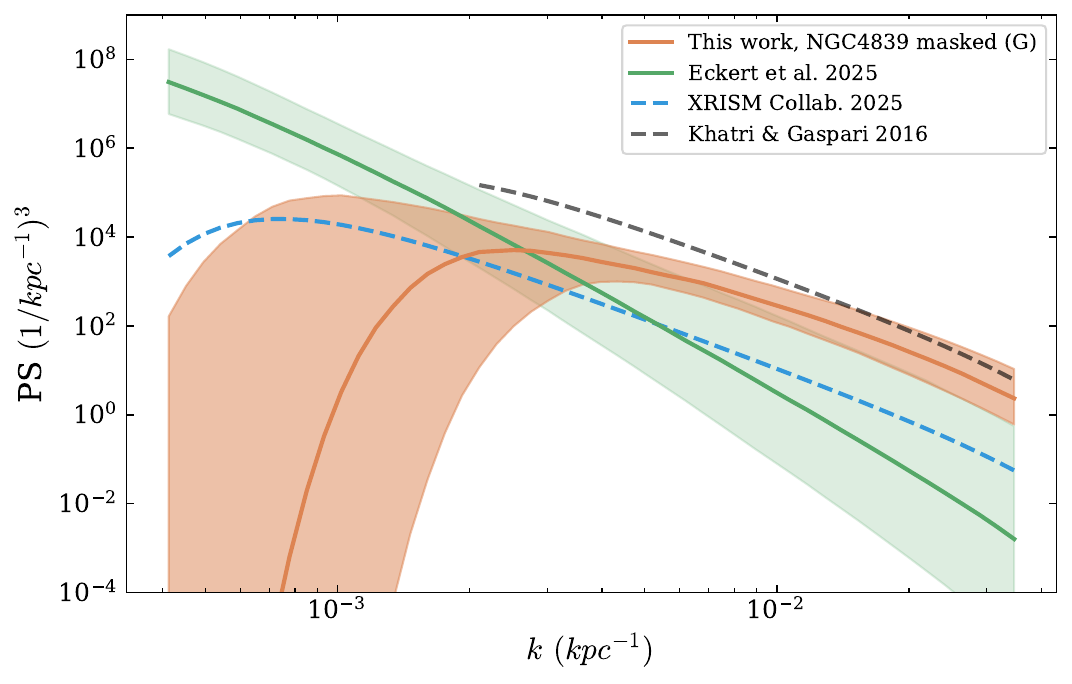}
    \caption{Reconstructed 3D power spectra using Eq. \ref{e:p3d} with their respective 68\% confidence envelopes. The spectrum of \cite{Eckert_2025} is computed from their original posterior distributions for the Fiducial case taking into account the uncertainties. The blue dashed curve represents the spectrum computed from \cite{xrismcollaboration2025xrismforecastcomacluster} (cf. Table \ref{t:results}). The black dashed line represent the inertial regime of the spectrum associated with the 3D Mach number of \cite{Khatri_2016} with a 11/3 slope.}
    \label{f:keyplot}
\end{figure}

The reconstructed 3D power spectrum of the pressure fluctuations,  $\mathcal{P}_{3D}$, is presented in Fig. \ref{f:keyplot}. We discuss in the next section the physical implications of our results in light of previous studies, while acknowledging potential limitations related to data quality and the assumptions underlying our analysis.

\renewcommand{\arraystretch}{1.6}
\begin{table*}[t]
\caption{Turbulent power spectrum parameters, with best fit results from our SBI analysis compared with those from relevant published works. The errors represent 68\% confidence intervals.}                 
\label{t:results}    
\centering                        
\normalsize
\begin{tabular}{c | c c c | c c c c}      
\hline\hline               
Studies & $\log(l_{inj})$  & $\alpha$ & $log(\sigma)$ & $l_{\text{inj}}$ & $\mathcal{M}_{3D, tot}$ &$\sigma_{v, \textbf{ }3D}$ & $P_{turb}/P_{tot}$ \\
         & (-)    & (-)   & (-) & (kpc) & (-) & (km/s)  & (-)     \\         
\hline                      

    Center fit + mask (U) & $-0.27^{+0.33}_{-0.21}$ &  $2.96^{+0.80}_{-0.35}$ & $-0.34^{+0.29}_{-0.17}$  & $530^{+790}_{-200}$ & $0.56^{+0.18}_{-0.06}$ & $\sim 350-1110$ & $0.14^{+0.09}_{-0.03}$ \\
    
    Center fit + mask (G) & $-0.30^{+0.23}_{-0.21}$ &  $3.50^{+0.50}_{-0.46}$ & $-0.26^{+0.22}_{-0.20}$  & $540^{+450}_{-200}$ & $0.60^{+0.13}_{-0.09}$ & $\sim 357-1095$ & $0.17^{+0.06}_{-0.04}$ \\
\hline                      

    Eckert Fiducial$^{(1)}$ & - &  $5.3^{+2.1}_{-1.0}$    & - & $2.2^{+2.0}_{-1.0}$ Mpc & $0.72^{+0.28}_{-0.22}$ & $\sim 1250$ & $0.10^{+0.08}_{-0.04}$\\

    XRISM Collab$^{(2)}$ & - & 11/3 & - & 1~Mpc (fixed) & $0.24^{+0.015}_{-0.015}$ & $350^{+42}_{-42}$$^{\text{ }(4)}$ & $0.031^{+0.004}_{-0.004}$ \\

    Khatri \& Gaspari 2016$^{(3)}$ & - & - & - & $\sim 500$ & $0.8^{+0.3}_{-0.3}$ & - &  $0.15-0.45$ \\
\hline                                  
\end{tabular}
\begin{tablenotes}
\small
\item (1) \citet{Eckert_2025}; (2) \citet{xrismcollaboration2025xrismforecastcomacluster}; (3) \citet{Khatri_2016}; (4) The reported value for $\sigma_{v, \textbf{ }3D}$ is directly converted with a $\sqrt{3}$ factor assuming isotropy.
\end{tablenotes}
\end{table*}

\section{Discussion}
\label{s:discu}

\subsection{Review of model assumptions}
\subsubsection{Impact of masking} 
\label{s:mas}

\begin{figure}[t]
    \centering
    \includegraphics[width=0.6\linewidth]{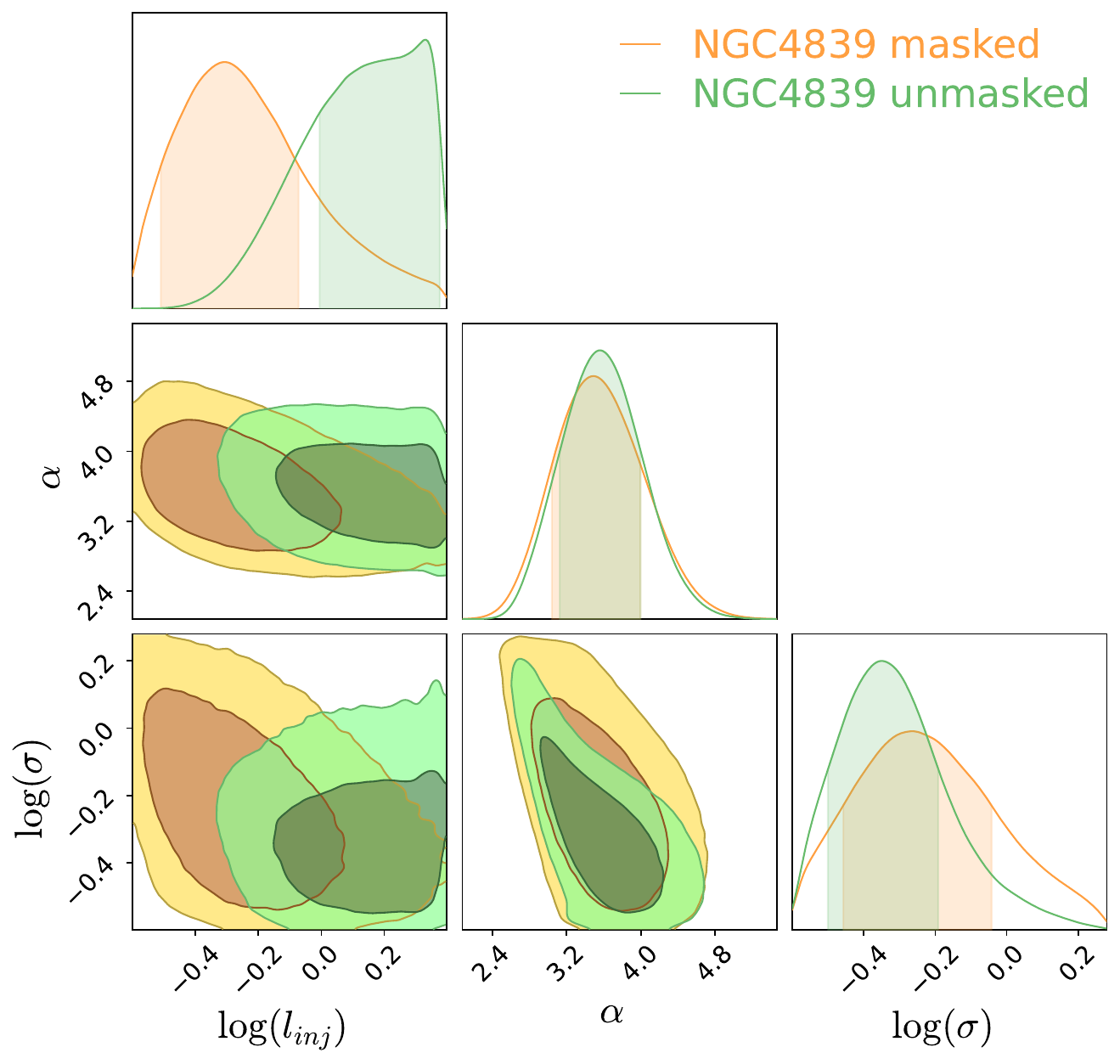}
    \caption{Posterior distribution for the three free parameters, comparing the reference case with and without NGC~4839 masked in the analysis. The prior on the slope $\alpha$ is Gaussian.}
    \label{f:maskimpact}
\end{figure}

Our results were obtained by masking the south-eastern region around the position of NGC~4839 for modelling the main cluster emission and in order to compute the 2D power spectrum of the Coma $y$-map fluctuations. Indeed, the perturbations induced by the bulk motions of this infalling group may bias the interpretation of the overall $y$-fluctuation statistics as turbulence. To assess the impact of this masking, we repeated our analysis without the pie-slice mask covering NGC~4839 and its infalling tail (see left panel of Fig.~\ref{f:szb}). We recomputed the 2D power spectrum, performed the simulation-based inference (SBI) training, and inferred the turbulent parameters. The derived posterior distributions, compared to the reference masked case, are shown in Fig.~\ref{f:maskimpact}.

The best-fit parameters obtained without masking are $\alpha = 3.56^{+0.44}_{-0.43}$, $\log(\sigma) = -0.35^{+0.16}_{-0.15}$, and $l_{\textrm{inj}} = 1330^{+1110}_{-460}$ kpc. The posterior distributions for all three free parameters remain  consistent within the $68\%$ confidence interval with respect to the masked case. However, the distribution of $l_{inj}$ is the most affected, exhibiting a significant shift towards higher values. This shift is consistent with the larger-scale fluctuations induced by NGC~4839's ram-pressure's tail, as seen in the right panel of Fig.~\ref{f:szb}, which likely artificially inflate the inferred injection scale when interpreted as a turbulent Gaussian random field.

This test highlights the necessity for caution when attributing the entire budget of $y$-fluctuations, hence pressure fluctuations, to turbulent motions alone. However, in our case this concern is mitigated by our investigation of the largest scales. Specifically, our fluctuation statistics were established over a broad area of $4R_{500} \times 4R_{500}$, ensuring that we capture the entirety of Coma's outer regions, where accretion driven turbulence is expected to dominate. Additionally, the moderate spatial resolution of \textit{Planck} works in our favour, as it washes out the potential contamination from smaller scale pressure fluctuations (e.g., due to sloshing) that are prevalent in the cluster's central regions.

\subsubsection{Impact of the centre position}
\label{s:cen}

To minimise biases in the derived $y$-fluctuation maps, we opted to leave the cluster centre as a free parameter when fitting the mean pressure model. This approach ensures that the residual fluctuations are not biased  by an arbitrarily fixed centre position. While a quantified investigation on the impact of the choice of model on the fluctuation statistics is provided by \citet{Dupourque2023b}, we focus here on the specific impact of the mean model centre position.

Our best-fit centre position deviates from the \planck\ PSZ2 catalogue coordinates \citep{PSZ2}, with offsets of $\Delta\alpha = 0.46$~arcmin and $\Delta\delta = 3.25$~arcmin. On our tangential grid, these correspond to pixel offsets of $0.49$ and $3.45$ pixels for $x_c$ and $y_c$, respectively. When we repeated our analysis using the PSZ2 catalogue centre instead of our best-fit position, the resulting mean model, fluctuation map, and 2D power spectrum exhibited noticeable differences compared to our reference case. The ensuing SBI analysis failed to converge properly in this scenario, highlighting the sensitivity of our method to the choice of centre.

To verify that our algorithm is not biased by pressure fluctuations artificially shifting the fitted centre, we conducted the following test: we generated $100$ simulated $y$-maps of the Coma cluster using as centre our best-fit position. For each simulated $y$-map that includes pressure fluctuations with $\log(\sigma) \in [-0.6, 0.28]$, we then refitted the elliptical mean model, including the centre position , and compared the recovered centre coordinates to the initial values. The distributions of the fitted centres  are consistent with the input positions: $\bar{x_c} = 102.1$ and $\bar{y_c} = 105.0$ with standard deviations of $\sigma_{x_c} = 1.0$~pixels and $\sigma_{y_c} = 0.8$~pixels.  This is smaller than the characteristic scales of the fluctuations we can probe in our framework. We recall that we exclude in our analysis all spatial  scales smaller than 2 times the \planck\ FWHM, that is $14 \textrm{arcmin} \approx 380$~kpc. This confirms that our method is not dominated by spurious shifts induced by local fluctuations  and it  demonstrates that our decision to fix the centre position during SBI training does not introduce any significant systematic errors. 

We also stress that the differences between the PSZ2 centre and our fitted position arise from the  differences in the methods from which they derive.  The PSZ2 cluster's centre positions are obtained from the cluster detection algorithms used to build the \planck\ cluster catalogues.  These assume a fixed gNFW pressure distribution (i.e., \citet{Arnaud2010} universal profile) in a full spherical assumption, quite different from our elliptical projection on the plane of the sky. Differences between the two approaches obviously increase with the level of perturbation of the dynamical state of the cluster, the presence of sloshing, etc.

\subsection{Robustness of the slope estimation}
\label{s:slope}

The constrained estimation of the turbulent cascade slope, $\alpha$, remains a persistent challenge in studies of the Coma cluster, as highlighted by recent analyses \citep{xrismcollaboration2025xrismforecastcomacluster,Eckert_2025,Zhang2026}. Our investigation identifies two primary factors contributing to this issue.

The slope parameter $\alpha$ characterises the energy transfer across scales in the turbulent cascade, from injection to dissipation. However, our analysis is constrained by \planck spatial resolution, from which we adopted a conservative lower scale limit of $2 \times \text{FWHM}\approx 380$~kpc. With a simulated dissipation scale of $\sim 1$~kpc and a retrieved injection scale of $\sim 550$~kpc, a significant portion of the cascade remains unresolved in the \planck\ observations. 
Furthermore, constraining the turbulent cascade slope $\alpha$ via our SBI analysis requires either multiple independent realisations of the turbulent field or a single system with sufficient spatial dynamic range to capture the full cascade.
Though we reach the largest scales with the \planck\ SZ observations, our analysis fails to meet these criteria due again to our cut-off at $\sim 380$~kpc, and the cluster's  dynamical state \citep{Zhang2026}, leading to a limited constraint on $\alpha$, as the inferred slope becomes sensitive to stochastic variations which introduce significant sample variance. We implemented the case of a fixed $\alpha = 11/3$, inferring only $\log(l_{inj})$ and $\log(\sigma)$. Whilst this setup does not significantly improve the constraints on the injection scale, it naturally reduces the uncertainties on $\log(\sigma)$. However, we found this case overly restrictive when compared to that of the Gaussian prior for the slope.

Simulations by \citet{Zhang2026} suggest that the merger induced dynamics in the Coma cluster may disrupt the turbulent cascade and explain the recent results on the gas velocity measurements  \citep{xrismcollaboration2025xrismforecastcomacluster}. In such scenarios, turbulence lacks sufficient time to fully develop. As such, the assumption that the velocity field can be considered as an homogeneous Gaussian random field is limited and too simplistic, potentially explaining the poor grasp and constraints on $\alpha$ in the data.

\subsection{Attempt to investigate turbulence with smaller apertures}

To study the scale dependence of turbulence in the Coma cluster, we performed our SBI analysis (with a Gaussian prior on the slope) on the $y$-map fluctuations using increasing circular apertures from 1$R_{500}$ to 2$R_{500}$ in radius, in steps of $0.2 R_{500}$.  This approach, similar to \citet{Khatri_2016}, probes how the recovered turbulence properties may vary with scale. The constraint on the slope remains unchanged, with posterior distributions consistent with the prior. The $\log(\sigma)$ inferences show larger uncertainties for larger apertures, while results remain consistent. The main differences are in $\log(l_{inj})$, with the posterior maximum values ranging from -0.01 ($1.4 R_\textit{500}$ aperture) to -0.30 (our $2 R_\textit{500}\times 2R_\textit{500}$,  Fig.~\ref{f:psz}), corresponding to physical injection scales of 1070~kpc and 540~kpc, respectively. All results remain consistent within their 68\% confidence intervals.
Reducing the size of the mask increases the sample variance at all spatial scales, as it naturally reduces the number of sampled modes and the number of spatially resolved elements. Conversely, using the full map increases noise contamination, thus lowering the overall signal of the 2D power spectrum. Furthermore, our assumption of a homogeneous Gaussian random field for the turbulent stochastic process may not hold over the entire field, as a transition in the injection scale of the fluctuation field has already been observed \citep{Dupourque2023b}. These cumulative factors can effectively lead to the aforementioned excursions in the reconstruction of the injection scale, as they can bias our modelling of the field structure. However, these effects mostly affect the injection scale and have no impact on the estimation of the Mach number, as it relies on the average statistic of the fluctuation field and not its characteristic structure.

\subsection{Understanding the turbulent processes in Coma}
\label{s:res_comparison}

This paper follows on from previous studies on the understanding of turbulent phenomena in the Coma cluster. The study of the pressure spectrum by \cite{Schuecker2004} has highlighted the presence of at least 10\% of the pressure in turbulent form. Analyses on small scales with \xmm and \chandra highlight that turbulence is not stopped at small scales of about $\sim 35$ kpc, which may be due to a smaller viscosity than expected \citep{Zhuravleva_2019}.

Recent studies, including this work, have constrained turbulence injection scales to large values such as $\approx 500$~kpc \citep{Khatri_2016}, $2.2^{+2.0}_{-1.0}$~Mpc \citep{Eckert_2025}, and $l_{inj} = 540^{+450}_{-200}$~kpc (this work), suggesting that turbulence in Coma's ICM is driven at large-scale. 

These high values for the retrieved injection scale reflect Coma's complex dynamical state, potentially influenced by its dual-BCG merger system \citep{Zhang2026} and the merging of NGC~4839, whose gravitational interactions could impact gas motions even beyond our masked regions.

As discussed in Section~\ref{s:slope}, both physical complexities and methodological limitations currently prevent strong constraints on the turbulent cascade slope. Whilst our results remain compatible with the Kolmogorov expectation ($\alpha=-11/3$), the broad posterior distributions shown in Figure~\ref{f:post} (for the uniform prior case) are consistent with other recent findings of steeper slopes \citep{xrismcollaboration2025xrismforecastcomacluster,Eckert_2025}, though these suffer from XRISM's limited spatial coverage of the Coma cluster. Altogether, the current analysis failed to converge to a clear constraint on the value of the turbulent cascade slope.

The characterisation of ICM turbulence through $\mathcal{M}_{3D}$ (directly linked to $\sigma_{v,\text{ }3D}$ and $P_{turb}/P_{tot}$) reveals systematic discrepancies across approaches. Our SZ fluctuation analysis, interpreting all observed fluctuations as the result of turbulent motion, yields $\mathcal{M}_{3D}=0.60^{+0.13}_{-0.09}$ (Table~\ref{t:results}), potentially overestimating true values. This contrasts with \citet{xrismcollaboration2025xrismforecastcomacluster}'s line-broadening measurement, $\mathcal{M}_{3D}=0.24\pm0.015$, directly due to line-of-sight velocity structures within two 9 arcmin$^2$ XRISM's pointings potentially highly affected by sample variance.
Conversely, two independent analyses align with our findings. \citet{Eckert_2025} report $\mathcal{M}_{3D,\text{tot}}=0.72^{+0.28}_{-0.22}$ from a power spectrum analysis using the same XRISM data (with an added  pointing) and a similar SBI approach to ours, whilst \citet{Khatri_2016}'s fluctuation analysis of the \planck\ SZ data gives an estimate of $\mathcal{M}_{3D}=0.8\pm0.3$. Though physically challenging, these two latter results remain statistically compatible at 68\% confidence with our findings, hence further tightening and strengthening the constraints obtained with \planck\ data.
These $\mathcal{M}_{3D}$ variations, directly tied to the power spectrum normalisation (Fig.~\ref{f:keyplot}), underscore the critical dependence of turbulence constraints on datasets, analytical methods as well as underlying assumptions \citep{Zhuravleva2023}. The convergence between \citet{Eckert_2025}, \citet{Khatri_2016} and our results, despite different datasets and methodologies, ultimately emphasise the robustness of our rigorous analytical framework and constraints on $\mathcal{M}_{3D}$, and derived values of $\sigma_{v,\text{ }3D}$  and $P_{turb}/P_{tot}$.

\begin{figure}[t]
    \centering
    \includegraphics[width=0.9\linewidth]{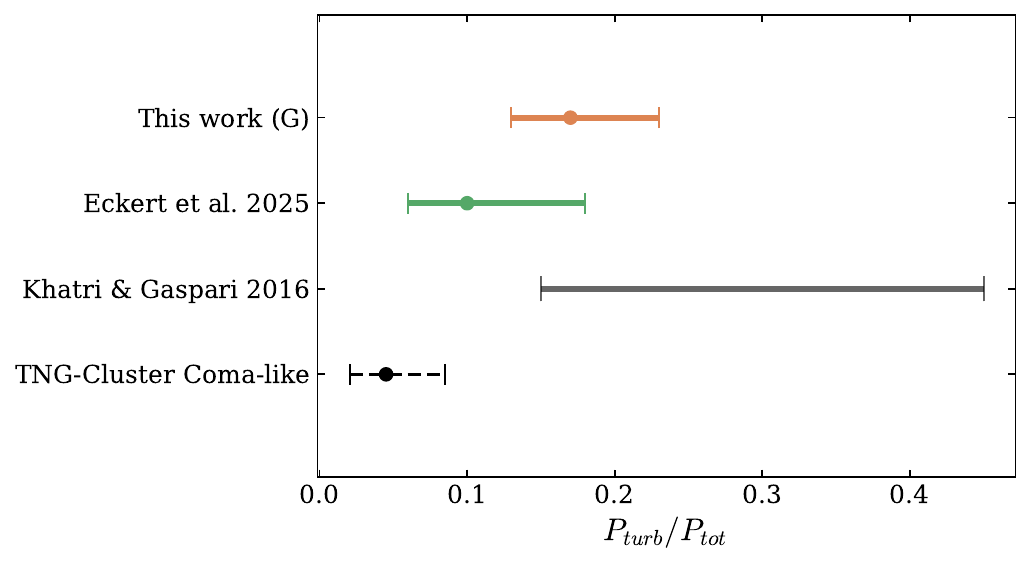}
    \caption{Constraints obtained on $P_{turb}/P_{tot}$ in recent Coma studies with similar data or methodology \citep[][respectively]{Khatri_2016, Eckert_2025}. We also display the prediction for Coma-like clusters used in \citet{xrismcollaboration2025xrismforecastcomacluster}, and derived from numerical simulations \citep{Nelson2024}.}
    \label{f:pressureratio}
\end{figure}

The turbulent to total pressure ratio (Eq.~\ref{eq:eturbetherm}) quantifies the hydrostatic mass bias when non-thermal pressure support is neglected, and provides an insight on the role of non-thermal processes (turbulence in our case)  in the virialisation of halos. Our derived value of $\sim17\%$ aligns with \citet{Eckert_2025}'s XRISM constraints (Fig.~\ref{f:pressureratio}) and remains consistent at 90\% confidence with the $\sim 8\%$ value derived from the X-ray fluctuation analysis within $0.6-1.0\,R_{500}$ over the X-COP sample \citep{Dupourque2023b}. It also agrees at 68\% confidence with the $14.2\%$ sample average from NIKA2/IRAM-30m SZ observations \citep{Adam2026}. However, the significantly lower $3.1\%$ constraint from Coma's XRISM line broadening \citep{xrismcollaboration2025xrismforecastcomacluster} stands in tension with these higher estimates, highlighting persistent discrepancies arising from  variations in datasets, spatial coverage, and methodological approaches.

\section{Conclusions and prospects}
\label{s:conclu}

In this paper, we presented a new SZ fluctuations analysis of the Coma cluster using \planck data. We developed a Simulation-Based Inference methodology to retrieve the turbulent parameters: $\log(l_{inj})$, $\alpha$, and $\log(\sigma)$, corresponding to the injection scale, spectral slope, and fluctuation amplitude, respectively.
This method is an adaptation of the one used for X-ray brightness fluctuations in \cite{Dupourque2023b} and also used for the analysis of direct measurements from XRISM in \cite{Eckert_2025} for the Simulation-Based Inference approach. The use of machine learning for the processing of Planck data on Coma enriches previous results. 

This study presents a novel analysis of Coma cluster SZ brightness fluctuations using \planck\ data. It implements an advanced SBI framework to constrain key  parameters of the turbulent power spectrum such as the injection scale, the power spectrum slope and the velocity dispersion. Our methodology builds upon and extends previous SBI applications to X-ray brightness fluctuations by \citet{Dupourque2023b}, representing the first machine learning-based analysis of \planck\ SZ data for Coma. Our rigorous and innovative approach significantly enhances the characterisation of ICM turbulence by combining the spatial coverage with the statistical power of neural posterior estimation, and accounting for a full error budget including the effect of sample variance. 

Our machine learning framework has delivered robust constraints on the parameters of our turbulence model when applied to the Coma cluster, yielding $l_{inj} = 540^{+450}_{-200}$ kpc, $\alpha = 3.50^{+0.50}_{-0.46}$, and $\log\sigma = -0.26^{+0.22}_{-0.20}$. Through the 3D power spectrum normalisation, we derived key turbulent properties: The 3D Mach number, $\mathcal{M}_{3D} = 0.60^{+0.13}_{-0.09}$ ; the velocity dispersion $\sigma_{v,\text{ }3D} = 357-1095$ km/s (depending on the speed of sound, function of the ICM gas temperature), and non-thermal pressure support induced by the turbulence $P_{turb}/P_{tot} = 0.17^{+0.06}_{-0.04}$. Our findings not only align with \citet{Khatri_2016}'s \planck\ SZ fluctuation analysis but also provide more robust and tighter constraints. Moreover, they exhibit excellent agreement with \citet{Eckert_2025}'s SBI-based power spectrum analysis of XRISM measurements, demonstrating remarkable cross-method consistency on very different datasets.

However, whilst our analysis provides some constraints on the turbulent cascade slope, important methodological caveats persist. Constraining $\alpha$ within our SBI framework proved to be difficult, mitigated by the adoption of an informative Gaussian prior for this parameter. This challenge mirrors recent findings \citep{Eckert_2025,xrismcollaboration2025xrismforecastcomacluster}, collectively suggesting either information limits in the current datasets or inherent shortcomings in turbulence modelling approaches.

Our analysis  provides unprecedented constraints on both the turbulent injection scale and the 3D Mach number. This study successfully adapts and implements the X-ray brightness fluctuation analysis framework to SZ observations, establishing a powerful methodological approach for turbulence characterisation. While Coma's exceptional brightness and angular extent made it an optimal and unique case study in SZ with \planck\ data, extending this methodology to more compact systems will require enhanced spatial resolution at millimetre wavelengths. For intermediate redshift clusters, dedicated high-resolution observations with NIKA2/IRAM-30m \citep[see e.g.][]{Adam2026} would be ideal, while for nearby clusters, combining ACT/SPT with \planck\ data to achieve $\sim1.5-2$~arcmin resolution represents the most promising path forward.

The next crucial advance will be to combine X-ray and SZ fluctuations analysis in a  self consistent approach over the cluster population, and correlate it to constraints from  ongoing high-resolution X-ray spectroscopic measurements from XRISM. This multi-probe approach will enable a systematic cross-validation of the properties of turbulence, providing the comprehensive observational constraints needed to  quantify the role of turbulence in shaping galaxy clusters.


\begin{acknowledgements}
We thank the referee for the useful review and the careful reading of our manuscript. We are grateful to Dominique Eckert for providing us with his raw SBI posterior distributions displayed in Fig. \ref{f:keyplot}.
This work benefited the support from CNRS/INSU and CNES (the French space agency). This work was also supported by the French government through the France 2030 investment plan managed by the National Research Agency (ANR), as part of the Initiative of Excellence of Université Côte d’Azur under reference number ANR-15-IDEX-01. We thank the contributors to the various open-source python packages such as \texttt{matplotlib} \citep{matplotlib}, \texttt{astropy} \citep{astropy}, \texttt{ChainConsumer} \citep{chainconsumer}, \texttt{sbi} \citep{sbi}, \texttt{jax} \citep{jax2018github}.
\end{acknowledgements}

%

\bibliographystyle{aa} 
\bibliography{ref}

\end{document}